\documentclass[prd,twocolumn,showpacs,amsmath,amssymb,floatfix]{revtex4}

\usepackage{graphicx}
\usepackage{dcolumn}
\usepackage{bm}

\newcommand{\ks}{$K_s$}

\newcommand{\pt}{$p_{T}$}
\newcommand{\neutralino}{$\chi^0_1$}

\begin{document}

\title{Search for neutral, long-lived particles decaying into two muons 
in \mbox{\boldmath $p\bar{p}$} collisions 
at \mbox{\boldmath $\sqrt{s}$} = 1.96 TeV 
}

%
\author{                                                                      
V.M.~Abazov,$^{36}$                                                           
B.~Abbott,$^{76}$                                                             
M.~Abolins,$^{66}$                                                            
B.S.~Acharya,$^{29}$                                                          
M.~Adams,$^{52}$                                                              
T.~Adams,$^{50}$                                                              
M.~Agelou,$^{18}$                                                             
J.-L.~Agram,$^{19}$                                                           
S.H.~Ahn,$^{31}$                                                              
M.~Ahsan,$^{60}$
% ...
\\                                                                            
\vskip 0.30cm                                                                 
\centerline{(D\O\ Collaboration)}                                             
\vskip 0.30cm                                                                 
}                                                                             
\affiliation{                                                                 
\centerline{$^{1}$Universidad de Buenos Aires, Buenos Aires, Argentina}       
\centerline{$^{2}$LAFEX, Centro Brasileiro de Pesquisas F{\'\i}sicas,         
                  Rio de Janeiro, Brazil}                                     
\centerline{$^{3}$Universidade do Estado do Rio de Janeiro,                   
                  Rio de Janeiro, Brazil}                                     
\centerline{$^{4}$Instituto de F\'{\i}sica Te\'orica, Universidade            
                  Estadual Paulista, S\~ao Paulo, Brazil}                     
\centerline{$^{5}$University of Alberta, Edmonton, Alberta, Canada,           
                  Simon Fraser University, Burnaby, British Columbia, Canada,}
\centerline{York University, Toronto, Ontario, Canada, and                    
                  McGill University, Montreal, Quebec, Canada}                
\centerline{$^{6}$Institute of High Energy Physics, Beijing,                  
                  People's Republic of China}                                 
\centerline{$^{7}$University of Science and Technology of China, Hefei,       
                  People's Republic of China}                                 
\centerline{$^{8}$Universidad de los Andes, Bogot\'{a}, Colombia}             
\centerline{$^{9}$Center for Particle Physics, Charles University,            
                  Prague, Czech Republic}                                     
\centerline{$^{10}$Czech Technical University, Prague, Czech Republic}        
\centerline{$^{11}$Center for Particle Physics, Institute of Physics,         
                   Academy of Sciences of the Czech Republic,                 
                   Prague, Czech Republic}                                    
\centerline{$^{12}$Universidad San Francisco de Quito, Quito, Ecuador}        
\centerline{$^{13}$Laboratoire de Physique Corpusculaire, IN2P3-CNRS,         
                   Universit\'e Blaise Pascal, Clermont-Ferrand, France}      
\centerline{$^{14}$Laboratoire de Physique Subatomique et de Cosmologie,      
                   IN2P3-CNRS, Universite de Grenoble 1, Grenoble, France}    
\centerline{$^{15}$CPPM, IN2P3-CNRS, Universit\'e de la M\'editerran\'ee,     
                   Marseille, France}                                         
\centerline{$^{16}$IN2P3-CNRS, Laboratoire de l'Acc\'el\'erateur              
                   Lin\'eaire, Orsay, France}                                 
\centerline{$^{17}$LPNHE, IN2P3-CNRS, Universit\'es Paris VI and VII,         
                   Paris, France}                                             
\centerline{$^{18}$DAPNIA/Service de Physique des Particules, CEA, Saclay,    
                   France}                                                    
\centerline{$^{19}$IPHC, IN2P3-CNRS, Universit\'e Louis Pasteur, Strasbourg,  
                    France, and Universit\'e de Haute Alsace,                 
                    Mulhouse, France}                                         
\centerline{$^{20}$Institut de Physique Nucl\'eaire de Lyon, IN2P3-CNRS,      
                   Universit\'e Claude Bernard, Villeurbanne, France}         
\centerline{$^{21}$III. Physikalisches Institut A, RWTH Aachen,               
                   Aachen, Germany}                                           
\centerline{$^{22}$Physikalisches Institut, Universit{\"a}t Bonn,             
                   Bonn, Germany}                                             
\centerline{$^{23}$Physikalisches Institut, Universit{\"a}t Freiburg,         
                   Freiburg, Germany}                                         
\centerline{$^{24}$Institut f{\"u}r Physik, Universit{\"a}t Mainz,            
                   Mainz, Germany}                                            
\centerline{$^{25}$Ludwig-Maximilians-Universit{\"a}t M{\"u}nchen,            
                   M{\"u}nchen, Germany}                                      
\centerline{$^{26}$Fachbereich Physik, University of Wuppertal,               
                   Wuppertal, Germany}                                        
\centerline{$^{27}$Panjab University, Chandigarh, India}                      
\centerline{$^{28}$Delhi University, Delhi, India}                            
\centerline{$^{29}$Tata Institute of Fundamental Research, Mumbai, India}     
\centerline{$^{30}$University College Dublin, Dublin, Ireland}                
\centerline{$^{31}$Korea Detector Laboratory, Korea University,               
                   Seoul, Korea}                                              
\centerline{$^{32}$SungKyunKwan University, Suwon, Korea}                     
\centerline{$^{33}$CINVESTAV, Mexico City, Mexico}                            
\centerline{$^{34}$FOM-Institute NIKHEF and University of                     
                   Amsterdam/NIKHEF, Amsterdam, The Netherlands}              
\centerline{$^{35}$Radboud University Nijmegen/NIKHEF, Nijmegen, The          
                  Netherlands}                                                
\centerline{$^{36}$Joint Institute for Nuclear Research, Dubna, Russia}       
\centerline{$^{37}$Institute for Theoretical and Experimental Physics,        
                   Moscow, Russia}                                            
\centerline{$^{38}$Moscow State University, Moscow, Russia}                   
\centerline{$^{39}$Institute for High Energy Physics, Protvino, Russia}       
\centerline{$^{40}$Petersburg Nuclear Physics Institute,                      
                   St. Petersburg, Russia}                                    
\centerline{$^{41}$Lund University, Lund, Sweden, Royal Institute of          
                   Technology and Stockholm University, Stockholm,            
                   Sweden, and}                                               
\centerline{Uppsala University, Uppsala, Sweden}                              
\centerline{$^{42}$Physik Institut der Universit{\"a}t Z{\"u}rich,            
                   Z{\"u}rich, Switzerland}                                   
\centerline{$^{43}$Lancaster University, Lancaster, United Kingdom}           
\centerline{$^{44}$Imperial College, London, United Kingdom}                  
\centerline{$^{45}$University of Manchester, Manchester, United Kingdom}      
\centerline{$^{46}$University of Arizona, Tucson, Arizona 85721, USA}         
\centerline{$^{47}$Lawrence Berkeley National Laboratory and University of    
                   California, Berkeley, California 94720, USA}               
\centerline{$^{48}$California State University, Fresno, California 93740, USA}
\centerline{$^{49}$University of California, Riverside, California 92521, USA}
\centerline{$^{50}$Florida State University, Tallahassee, Florida 32306, USA} 
\centerline{$^{51}$Fermi National Accelerator Laboratory,                     
            Batavia, Illinois 60510, USA}                                     
\centerline{$^{52}$University of Illinois at Chicago,                         
            Chicago, Illinois 60607, USA}                                     
\centerline{$^{53}$Northern Illinois University, DeKalb, Illinois 60115, USA} 
\centerline{$^{54}$Northwestern University, Evanston, Illinois 60208, USA}    
\centerline{$^{55}$Indiana University, Bloomington, Indiana 47405, USA}       
\centerline{$^{56}$University of Notre Dame, Notre Dame, Indiana 46556, USA}  
\centerline{$^{57}$Purdue University Calumet, Hammond, Indiana 46323, USA}    
\centerline{$^{58}$Iowa State University, Ames, Iowa 50011, USA}              
\centerline{$^{59}$University of Kansas, Lawrence, Kansas 66045, USA}         
\centerline{$^{60}$Kansas State University, Manhattan, Kansas 66506, USA}     
\centerline{$^{61}$Louisiana Tech University, Ruston, Louisiana 71272, USA}   
\centerline{$^{62}$University of Maryland, College Park, Maryland 20742, USA} 
\centerline{$^{63}$Boston University, Boston, Massachusetts 02215, USA}       
\centerline{$^{64}$Northeastern University, Boston, Massachusetts 02115, USA} 
\centerline{$^{65}$University of Michigan, Ann Arbor, Michigan 48109, USA}    
\centerline{$^{66}$Michigan State University,                                 
            East Lansing, Michigan 48824, USA}                                
\centerline{$^{67}$University of Mississippi,                                 
            University, Mississippi 38677, USA}                               
\centerline{$^{68}$University of Nebraska, Lincoln, Nebraska 68588, USA}      
\centerline{$^{69}$Princeton University, Princeton, New Jersey 08544, USA}    
\centerline{$^{70}$State University of New York, Buffalo, New York 14260, USA}
\centerline{$^{71}$Columbia University, New York, New York 10027, USA}        
\centerline{$^{72}$University of Rochester, Rochester, New York 14627, USA}   
\centerline{$^{73}$State University of New York,                              
            Stony Brook, New York 11794, USA}                                 
\centerline{$^{74}$Brookhaven National Laboratory, Upton, New York 11973, USA}
\centerline{$^{75}$Langston University, Langston, Oklahoma 73050, USA}        
\centerline{$^{76}$University of Oklahoma, Norman, Oklahoma 73019, USA}       
\centerline{$^{77}$Oklahoma State University, Stillwater, Oklahoma 74078, USA}
\centerline{$^{78}$Brown University, Providence, Rhode Island 02912, USA}     
\centerline{$^{79}$University of Texas, Arlington, Texas 76019, USA}          
\centerline{$^{80}$Southern Methodist University, Dallas, Texas 75275, USA}   
\centerline{$^{81}$Rice University, Houston, Texas 77005, USA}                
\centerline{$^{82}$University of Virginia, Charlottesville,                   
            Virginia 22901, USA}                                              
\centerline{$^{83}$University of Washington, Seattle, Washington 98195, USA}  
}                                                                             

\date{\today}
           
\begin{abstract}
We present a search for a neutral particle, pair-produced in $p\bar{p}$
collisions at $\sqrt{s}$=1.96 TeV, which decays into two muons 
and lives long enough
to travel at least 5 cm before decaying.  The analysis uses 
$\approx$380 pb$^{-1}$ of data recorded with the D0 detector.
The background is estimated to be about one event.
No candidates are observed, and limits are set on the pair
production cross section times branching fraction into dimuons + $X$
for such particles.  For a mass of 10 GeV and lifetime of 
$4 \times 10^{-11}$~s, we exclude values greater than
0.14 pb (95\% C.L.).  These results are used to limit the 
interpretation of NuTeV's excess of di-muon events.
\end{abstract}


\maketitle

\newpage

Several models including supersymmetry with $R$-parity
violation~\cite{bib:umssm,bib:rparity} and 
hidden valley theories~\cite{bib:strassler} predict the existence of neutral,
long-lived particles that give rise to a distinctive signature of two 
leptons arising from a highly displaced vertex.
The Fermilab neutrino experiment NuTeV observed an excess of 
di-muon events that could be interpreted as such a 
signal~\cite{bib:nutev3events,bib:luibo,bib:bbinterpret}.
Experiments at the CERN $e^+ e^-$ Collider (LEP) have looked for 
short-lived neutralino and chargino 
decays~\cite{bib:lepneutsearch} and longer-lived charged 
particles~\cite{bib:delphicmsp}, but did not search for this
topology.

In this Letter we present a search for a light, neutral,
long-lived particle ($N_{LL}^0$) pair-produced
in $p\bar{p}$ collisions at $\sqrt{s}$ = 1.96 TeV and recorded
with the D0 detector, using 380 pb$^{-1}$ of data from Run II 
of the Fermilab Tevatron Collider.  The final state
under study is the decay of an $N_{LL}^0$  
into two muons and possibly a neutrino
after the $N_{LL}^0$ has traveled at least 5 cm.
The particle is assumed to have a mass as low as several GeV.  The
analysis reported here explores a region of 
phase space previously unexplored by collider
experiments.  

\begin{figure}
 \begin{minipage}{4.1cm}
  \includegraphics[height=4.1cm,angle=270]{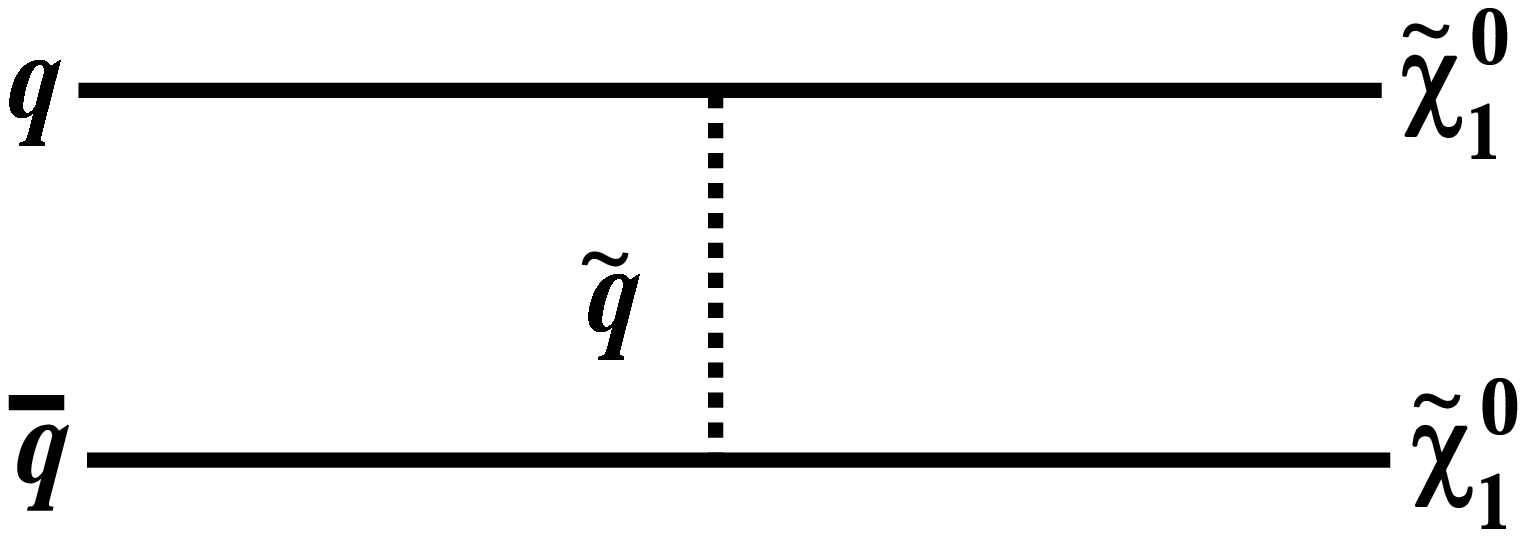}
 \end{minipage}
 \begin{minipage}{4.1cm}
  \includegraphics[height=4.1cm,angle=270]{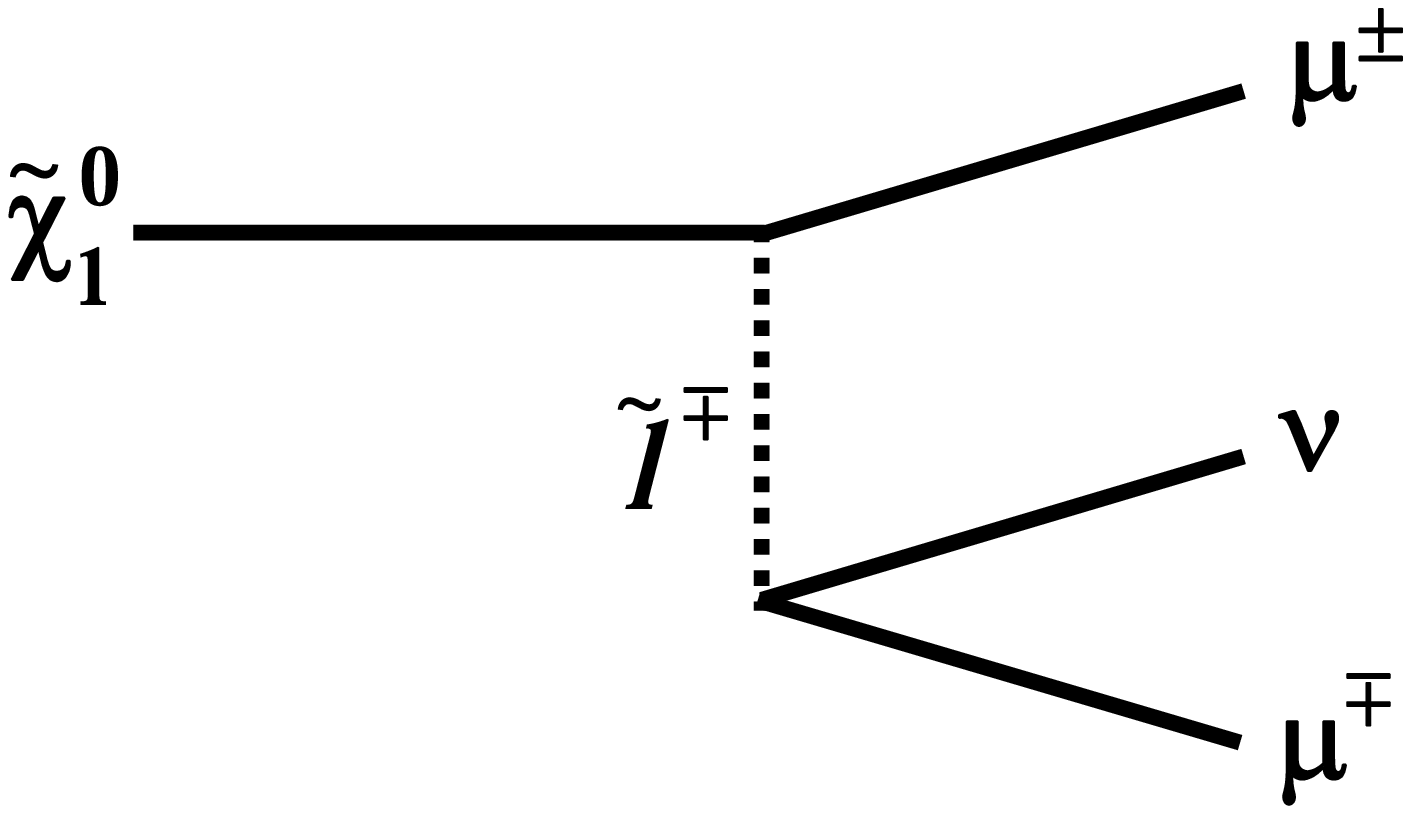}
 \end{minipage}
 \caption{Feynman diagrams for pair production (left) and decay (right) of 
   a neutral particle, in this case neutralinos with $R$-parity
   violation.  \label{fig:neutfeyn}}
\end{figure}

We use $R$-parity violating (RPV) decays of neutralinos (\neutralino) 
to $\mu^+\mu^-\nu$ (Fig.\ \ref{fig:neutfeyn})
as a benchmark model and to determine signal efficiency.  
Here the RPV couplings are expected to be small and lead to long 
lifetimes~\cite{bib:barbier}.
The results are applicable to any pair-produced neutral
particle with similar kinematics.

The D0 detector consists of a central-tracking system, a 
liquid-argon and uranium calorimeter, and an outer muon 
system~\cite{run2det}. 
Each of these is used in this analysis, with an emphasis
on the muon system for particle identification and on the tracking
system for momentum measurement and vertexing.  

The central tracker consists of a
silicon microstrip tracker (SMT) and a central fiber tracker (CFT),
both located within a 2~T superconducting solenoidal magnet.  
It is optimized for tracking and vertexing at 
pseudorapidities $|\eta|<3$ and $|\eta|<2.5$, respectively, where
$\eta$ = $-\ln[\tan(\theta/2)]$ and $\theta$ is the polar angle
with respect to the proton beam direction.
The CFT has 8 axial and 8 stereo layers with an
innermost (outermost) radius of 20 (52) cm.
The calorimeter consists of a 
central section (CC) covering $|\eta| \leq 1.1$, 
and two end calorimeters (EC) that extend coverage
to $|\eta|\approx 4.2$, with all three housed in separate
cryostats~\cite{run1det}. The outer muon system, at $|\eta|<2$,
consists of a layer of tracking detectors and scintillation trigger
counters in front of 1.8~T iron toroids, followed by two similar layers
after the toroids~\cite{run2muon}.

We use the volume inside the CFT inner radius as a
decay region.  This allows the full CFT and muon systems to be
used for detection of decay products, ensuring robust track
reconstruction and muon identification.  
Events are required to pass a di-muon trigger.

The strategy is to identify events with at least two opposite-sign,
isolated muons, 
defined as hits in the muon system matched to a track in the CFT.  
Each pair is fit to a vertex. The signal sample uses events with
muon vertices that are displaced more than 5 cm (in the plane transverse
to the beamline) from the primary vertex.  To characterize the displacement,
we define the {\em vertex radius} 
\begin{equation}
 r = \sqrt{(X - X_{PV})^2 + (Y - Y_{PV})^2}
 \label{eq:radius}
\end{equation}
where $X$,$Y$ are the $x$,$y$ positions of the
fit di-muon vertex and $X_{PV}$,$Y_{PV}$ are the $x$,$y$ positions
of the primary vertex (PV).
D0 uses a right-handed coordinate system with the positive $z$-axis 
defined by the proton direction and positive $y$-axis pointed
upward.

Studies of \ks\ mesons are performed to test the reconstruction
efficiency for 
highly displaced vertices.  We search for \ks\ mesons in 
data and Monte Carlo (MC) by fitting track pairs to a common vertex
and selecting those with an invariant mass around the \ks\ peak.
We are able to observe decay lengths greater than 20 cm and demonstrate 
that the data and MC follow the same radial dependence.  The efficiency
varies by 30\% in the range $r$=5--20 cm.

Selection criteria are chosen to minimize background while
maintaining signal efficiency.  
All possible primary and secondary vertices are determined
using tracks, except those associated with muons.  
The hard scatter vertex is determined by clustering tracks into
seed vertices by a Kalman filter algorithm~\cite{bib:kalman}.
A probability function based on the $p_T$ of tracks attached to
each vertex is used to rank the likelihood that
it comes from a minimum bias interaction.  
The primary vertex is the one with the lowest probability.
The PV is
required to be within 0.3 cm of the beamline in $x$ and $y$ and
within 60 cm of the detector center in $z$.  

We require two muons which have hits in each of the three layers
of the muon system, are
matched to a track in the central tracker, have a good track fit,
at least 14 CFT hits associated with the track,
transverse momentum $> 10$ GeV, and are isolated.
Two methods are used to define isolation.
First, the direction of the muon is projected to the calorimeter and
the transverse energy in all cells within an annular cone 
0.1 $< R <$ 0.4 is summed (calorimeter isolation), where 
$R =  \sqrt{(\Delta \phi)^2 + (\Delta \eta)^2}$.  Second, the
transverse momentum of all tracks within a cone of $R < 0.5$ 
(except for the muon tracks) is summed (track isolation).  Both
the calorimeter and track isolations are required to be less than
2.5 GeV.
Cosmic ray muons are rejected by requiring the time measured by the
muon scintillator counters to be that expected
for a particle produced at the nominal beam crossing time.
To enhance the signal,
both muons must have a distance of closest approach (DCA) of 
greater than 0.01 cm in the $x-y$ plane and more than 0.1 cm along
the $z$-axis from any vertices.
The two muons must have an opening angle less than
0.5 radians and have opposite charge.  

All pairs of muons passing the above quality cuts are fit to a 
common vertex requiring a $\chi^2/N_{dof} < 4$.  
The radial distance between the di-muon vertex and the primary vertex 
must be six times the resolution of the di-muon vertex measurement 
and be between 5 and 20 cm.
This defines our signal region.

We use data to estimate the background for this
search. By allowing events to pass or fail
two different selection criteria (the DCA and vertex radius
cuts) we define four regions.  For the DCA cut,
we require either: (1) one track to pass the DCA cut and one to
fail it, or (2) both tracks to pass the DCA cut.  For the vertex
radius we define two regions:  (A) 0.3 $<$ $r$ $<$ 5 cm, or
(B) 5 $<$ $r$ $<$ 20 cm.  This defines Samples 1A, 2A, 1B, and 2B.

We observe four events in Sample 1A, one event in Sample 1B, and  
three events in Sample 2A.  
Assuming no correlation between selection criteria
and no signal, the ratio of the number of events in region to 2B to the
number in 1B should equal the ratio of the number of events in region 2A
to the number in 1A.  This can be re-expressed to give an 
estimate of the background in the signal sample (Sample 2B):
\begin{equation}
 N_{\text{bkgd}} = \frac{\rm Sample~2A}{\rm Sample~1A} \times {\rm Sample~1B}. 
    = 0.75 \pm 1.1~{\rm events}
 \label{eq:backgnd}
\end{equation}

Due to bias from a correlation
between the vertex radius and DCA criteria, we perform tests 
of this method and
assign a systematic uncertainty using several additional samples.
The spread in these results is used to assign a systematic
uncertainty to account for the correlation between the vertex radius
and DCA cut. Thus, we estimate the background
in the signal region to be 0.75 $\pm$ 1.1 (stat) $\pm$ 1.1 (syst) events.

\begin{figure}
 \includegraphics[width=8cm]{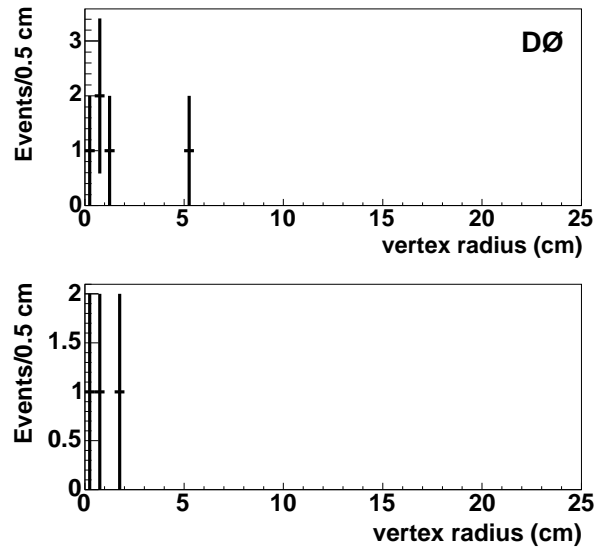}
 \caption{Distribution of the vertex radius for events where one muon passes
   the DCA criteria and the second fails it (top) and where both muons pass
   the DCA criteria (bottom).
   \label{fig:rdist}}
\end{figure}

Figure~\ref{fig:rdist} shows the vertex radius distribution for events
where one or both muons pass the DCA criteria.
Examination of the signal region yields 0 events passing all
criteria.  Therefore we set a limit on the cross section as a
function of lifetime.  
The lifetime dependence is calculated based on the fraction of 
events, $f$, which decay within our signal region.

Signal MC events are generated using
{\textsc{SUSYGEN}}~\cite{bib:susygen} and an
unconstrained minimal supersymmetric model with $R$-parity
violation~\cite{bib:luibo,bib:umssm,bib:rparity} using
CTEQ5L parton distribution functions (PDFs)~\cite{bib:cteq5}.  
The following parameters
are used: $\tan \beta = 10$, $\mu = -5000$, 
$M_2 = 200$ GeV, $M_3 = 400$ GeV, $M_{\text{squark}} = 300$ GeV, 
$M_{\text{slepton}}=M_{\text{snu}}=M_{\text{sbottom}}=M_{\text{stop}} 
= 1500$ GeV. 
The \neutralino\ mass is about equal to the $M_1$ parameter.  Similar
sets are generated with $M_1$ = 3, 5, 8, 10, 15, 20, 30, and 40 GeV yielding 
pair production cross sections in the range 
0.025--0.013 pb. 
While the parameters are different than 
those used in other 
neutralino searches~\cite{bib:lepneutsearch}, they are chosen to give a model 
that provides a particle consistent with the NuTeV result and does not 
violate limits from LEP searches (including the $Z$ boson invisible width).
The lifetime is
determined primarily by the slepton mass and the
$\lambda_{122}$ parameter, however, in the detector simulation we
choose to ignore the lifetime and force exactly one of the two \neutralino s
to decay within a cylinder of radius 25 cm.  The vertex is selected along
the \neutralino\ trajectory such that the radius distribution is flat over
the range 0--25 cm.  The other \neutralino\ is required to escape
the detector.  The dependence of the acceptance on the lifetime is accounted
for in the interpretation of the final result.  The average 
\neutralino\ transverse momentum (\pt) is $\approx$85 GeV.

Our uncertainty estimate on the luminosity times signal acceptance is
summarized in Table~\ref{tab:mclimits}.  The MC acceptance uncertainty
is statistical.
Tracking, isolation, and muon reconstruction data/MC corrections are 
estimated using the $Z$ boson mass peak, yielding 0.72 $\pm$ 0.07.
The vertex reconstruction data/MC is found using $K_s$ events
(0.92 $\pm$ 0.14).  A PDF uncertainty on the signal efficiency
of $\pm$4\% is assigned using the CTEQ6.1M PDF set~\cite{bib:cteq6}.

\begin{table}
 \caption{Acceptance, error, and limits for the MC signal points.
          The lifetime acceptance
          is the factor by which the limit is adjusted due to the
          fraction of events which decay within the 5--20 cm region and is
          given for a lifetime of $4 \times 10^{-11}$ s.  The
          luminosity $\times$ acceptance includes the MC signal acceptance,
          the trigger efficiency, the data/MC correction factors, and
          the luminosity.
          The limits are given for the same lifetime.
          \label{tab:mclimits} }
 \begin{center}
  \begin{tabular}{ccccc} \hline \hline
     & & & Luminosity & 95\% C.L. \\ 
    $M({\chi_0^1})$ & Monte Carlo & Lifetime & $\times$ acceptance & limit \\ 
    (GeV) & acceptance & acceptance & (pb$^{-1}$) & (pb) \\ \hline 
   3 & 0.095 $\pm$ 0.005 & 0.51 & 23.9 $\pm$ 4.8 & 0.28  \\
   5 & 0.114 $\pm$ 0.005 & 0.61 & 28.7 $\pm$ 5.8 & 0.19  \\
   8 & 0.141 $\pm$ 0.006 & 0.67 & 35.5 $\pm$ 7.1 & 0.14  \\  
  10 & 0.136 $\pm$ 0.006 & 0.68 & 34.3 $\pm$ 6.8 & 0.14  \\ 
  15 & 0.139 $\pm$ 0.006 & 0.65 & 35.1 $\pm$ 7.0 & 0.15  \\ 
  20 & 0.130 $\pm$ 0.005 & 0.62 & 32.8 $\pm$ 6.5 & 0.17  \\ 
  30 & 0.099 $\pm$ 0.003 & 0.55 & 24.8 $\pm$ 4.9 & 0.25  \\ 
  40 & 0.079 $\pm$ 0.004 & 0.48 & 20.0 $\pm$ 4.1 & 0.35  \\ \hline \hline
  \end{tabular}
 \end{center}
\end{table}

\begin{figure}
 \centerline{\includegraphics[width=8.9cm]{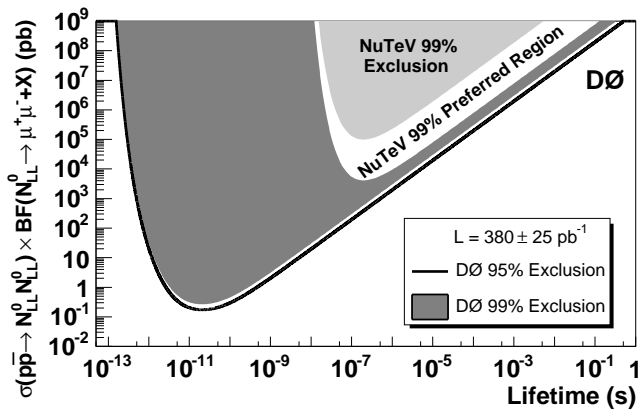}}
 \caption{Limit on cross section $\times$ branching fraction for the
    pair-production of neutral, long-lived particles as a function of
    lifetime.  The dark gray area and above represents the 
    D0 99\% C.L. limit for 
    the 5 GeV mass point.  The solid line shows the D0 95\% C.L. limit.
    The light gray 
    region represents the NuTeV 99\% C.L. exclusion~\cite{bib:nutev3events}
    converted to a
    $p\bar{p}$ cross section at $\sqrt{s} = 1960$ GeV.
    The white region represents a 99\% C.L. preferred region
    given the three events from NuTeV.
    \label{fig:expcompare}}
\end{figure}

\begin{figure}
 \includegraphics[width=8.6cm]{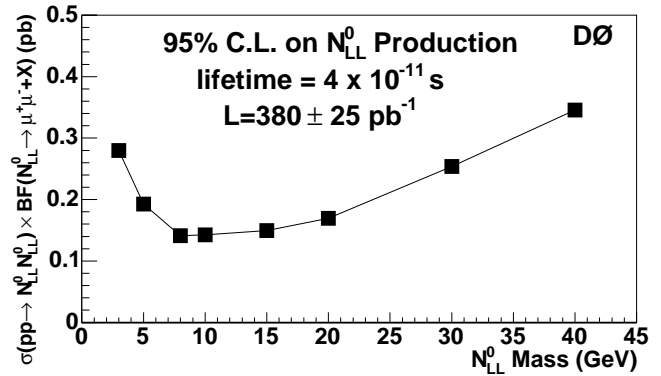}
 \caption{Limits on $N_{LL}^0$ pair production as a function of its mass.  
    The limit is for a lifetime of $4 \times 10^{-11} s$.
    \label{fig:masslimit}}
\end{figure}

These event numbers, efficiencies, acceptances and uncertainties
are combined to set a 95\% (99\%) confidence level limit on
the cross section $\sigma$($p\bar{p}\rightarrow N_{LL}^0N_{LL}^0X$) times
branching fraction BF($N_{LL}^0\rightarrow\mu^+\mu^- + X$)
as a function of the lifetime (Fig.\ \ref{fig:expcompare}),
using a Bayesian technique~\cite{bib:limit} and
assuming zero background.  The limit for a 10 GeV $N_{LL}^0$ with a
lifetime of $4 \times 10^{-11}$ s is 0.14 pb (95\% C.L.).
Figure~\ref{fig:masslimit} shows how the D0 limit varies with
mass at a lifetime of $4 \times 10^{-11}$ s.

In order to compare with D0, we convert the NuTeV result from $pp$ production 
at $\sqrt{s} = 38$ GeV to $p\bar{p}$ production at $\sqrt{s} = 1960$ GeV
using the ratio of cross sections for SUSY neutralino pair
production calculated with the parameters from our 5 GeV signal simulation.
The NuTeV lifetime is converted from kilometers to seconds
assuming an average momentum (along the neutrino beam direction) of 121 GeV.
Given that NuTeV observed three events~\cite{bib:nutev3events},
a preferred region is found using 
the ratio of the 99\% CL lower and upper limits on three events 
determined using a Feldman-Cousins approach~\cite{bib:feldman}.
We improve on the NuTeV limit by several orders of
magnitude at long lifetimes and add coverage at lower lifetimes.
Our limit excludes the interpretation of the NuTeV excess as 
arising from any model with similar $N_{LL}^0$ production cross sections
and kinematics.

To summarize, we have presented an analysis sensitive to neutral, long-lived
particles decaying to $\mu\mu + X$ using a new technique that
expands the capabilities of the D0 experiment.  
The background is estimated
to be 0.75 $\pm$ 1.1 (stat) $\pm$ 1.1 (syst) events.  
The signal region contains 0 events and a limit is set.
The 95\% CL limit for a mass of 10 GeV and
lifetime of $4 \times 10^{-11}$ s is 0.14 pb.  
This result excludes an interpretation of the NuTeV excess of dimuon
events in a large class of models.

%
We thank the staffs at Fermilab and collaborating institutions, 
and acknowledge support from the 
DOE and NSF (USA);
CEA and CNRS/IN2P3 (France);
FASI, Rosatom and RFBR (Russia);
CAPES, CNPq, FAPERJ, FAPESP and FUNDUNESP (Brazil);
DAE and DST (India);
Colciencias (Colombia);
CONACyT (Mexico);
KRF and KOSEF (Korea);
CONICET and UBACyT (Argentina);
FOM (The Netherlands);
PPARC (United Kingdom);
MSMT (Czech Republic);
CRC Program, CFI, NSERC and WestGrid Project (Canada);
BMBF and DFG (Germany);
SFI (Ireland);
The Swedish Research Council (Sweden);
Research Corporation;
Alexander von Humboldt Foundation;
and the Marie Curie Program.

\end{document}